\documentclass{article}

\usepackage{arxiv}

\usepackage[utf8]{inputenc} % allow utf-8 input
\usepackage[T1]{fontenc}    % use 8-bit T1 fonts
\usepackage{hyperref}       % hyperlinks
\usepackage{url}            % simple URL typesetting
\usepackage{booktabs}       % professional-quality tables
\usepackage{amsfonts}       % blackboard math symbols
\usepackage{nicefrac}       % compact symbols for 1/2, etc.
\usepackage{microtype}      % microtypography
\usepackage{lipsum}
\usepackage{graphicx}
\graphicspath{ {./images/} }

\title{Partnering with AI: A Pedagogical Feedback System for LLM Integration into Programming Education}

\author{
 Niklas Scholz \\
  Saarland Informatics Campus \\
  Saarland University \\
  Saarbrücken, Germany \\
  \texttt{nisc00019@stud.uni-saarland.de}
  \And
  Manh Hung Nguyen \\
  Max Planck Institute for Software Systems\\
  Saarbrücken, Germany\\
  \texttt{manguyen@mpi-sws.org} \\
  \And
  Adish Singla \\
  Max Planck Institute for Software Systems\\
  Saarbrücken, Germany\\
  \texttt{adishs@mpi-sws.org} \\
  \And 
  Tomohiro Nagashima \\
  Saarland Informatics Campus \\
  Saarland University \\
  Saarbrücken, Germany \\
  \texttt{nagashima@cs.uni-saarland.de}
}

\begin{document}
\maketitle
\begin{abstract}
Feedback is one of the most crucial components to facilitate effective learning. With the rise of large language models (LLMs) in recent years, research in programming education has increasingly focused on automated feedback generation to help teachers provide timely support to every student. However, prior studies often overlook key pedagogical principles, such as mastery and progress adaptation, that shape effective feedback strategies. This paper introduces a novel pedagogical framework for LLM-driven feedback generation derived from established feedback models and local insights from secondary school teachers.
To evaluate this framework, we implemented a web-based application for Python programming with LLM-based feedback that follows the framework and conducted a mixed-method evaluation with eight secondary-school computer science teachers. Our findings suggest that teachers consider that, when aligned with the framework, LLMs can effectively support students and even outperform human teachers in certain scenarios through instant and precise feedback. However, we also found several limitations, such as its inability to adapt feedback to dynamic classroom contexts. Such a limitation highlights the need to complement LLM-generated feedback with human expertise to ensure effective student learning. This work demonstrates an effective way to use LLMs for feedback while adhering to pedagogical standards and highlights important considerations for future systems.
\end{abstract}

% keywords can be removed

\section{Introduction}
Feedback is a fundamental strategy for supporting effective learning \cite{hattie_power_2007}, resulting in extensive research on effective strategies over the past decades. Prior work has developed various feedback frameworks, focusing on general principles of effective feedback \cite{nicol_formative_2006} or the need to tailor feedback to individual learners based on characteristics, such as mastery status \cite{mason_providing_2001,narciss_how_2004}. 

Recent advances in Large Language Models (LLMs) have significantly expanded the capabilities of adaptive feedback systems because of their ability to individually respond to open-ended text input from students in real time. Existing works leveraging LLMs with a particular focus on programming education, however, tend to focus on their technical aspects \cite{birillo_one_2024,guo_using_2024,lee_teachers_2024,lohr_youre_2025,phung_automating_2024,roest_next-step_2024}. Importantly, the majority of these recent works aims to streamline the process of giving feedback by having a single, simple prompt for all students, and therefore fails to incorporate adaptive pedagogical principles \cite{ma_how_2024,phung_automating_2024}. Since the effectiveness of feedback is closely tied to pedagogical principles \cite{nicol_formative_2006}, it is crucial to understand the specific capabilities and limitations of LLM-generated feedback when it comes to generating feedback aligned with such adaptive pedagogical principles. Hence, our work aims to answer the following design/research questions: 
\begin{enumerate}
    \item How might we leverage LLMs to design a feedback system that provides pedagogically-sound adaptive feedback in programming education?
\item To what extent does LLM-generated feedback compare to human teachers in terms of its pedagogical soundness?
\end{enumerate}
\section{Background: Automated Feedback Generation}

As the classroom size increases, teachers will find it increasingly hard to provide timely feedback and help to all their students. Past research has developed automated, adaptive feedback generation systems, many of which were rule-based systems employing techniques such as model tracing or, particularly in the programming domain, static and dynamic code analysis \cite{keuning_systematic_2018}. These systems, while effective to some extent, do not fully individualize feedback messages; instead, feedback messages are selected from a set of pre-determined answers based on current states of the learner \cite{silva_adaptive_2019}.

The rise of large language models (LLMs) in recent years has opened up new opportunities to enhance education across a range of tasks \cite{DBLP:journals/corr/abs-2402-01580}, including the development of conversational tutoring systems \cite{DBLP:conf/aied/SchmuckerXAM24}, student knowledge modeling \cite{DBLP:conf/edm/NguyenTS24}, and, in particular, automated feedback generation \cite{phung_automating_2024}. LLMs offer cost-effective ways to provide more individualized, elaborated responses to each student in real time. Hence, they have been used for feedback generation across domains, particularly in programming education. For instance, Phung et al. \cite{phung_automating_2024} developed a three-stage system using GPT-4 to generate single-sentence hints to address one of the bugs in student code and GPT-3.5 student agents for hint validation.
Birillo et al.  \cite{birillo_one_2024} combined static code analysis for extracting errors with LLMs to generate a next-step hint that significantly improved the effectiveness of feedback. Many automated feedback systems, however, do not incorporate any specific adaptation mechanisms within prompts or instruct LLMs to follow specific pedagogical principles but rather give simple instructions that do not adapt to learners' status (e.g., performance) \cite{birillo_one_2024,lee_teachers_2024,phung_automating_2024}. 

We argue that it is critical to design LLM-based feedback so that it aligns with pedagogical principles to effectively support student learning. Indeed, Lee and Song \cite{lee_teachers_2024} evaluated their system for programming education together with teachers and students, finding that the lack of pedagogical principles in the prompts made teachers feel skeptical about the LLM responses. To the best of our knowledge, this gap in exploring how to instruct LLMs to provide pedagogically sound feedback has not been explored. While there exist few studies considering different feedback strategies and pedagogical principles in their evaluation \cite{roest_next-step_2024} or exploring how to explicitly instruct LLMs to provide different types of feedback \cite{lohr_youre_2025}, those studies only show the potential of LLMs when it comes to effective feedback generation but fall short of offering practical suggestions for systems that adapt feedback strategies to diverse learner profiles \cite{zhang_students_2024}. 

\section{The Development and Implementation of a Pedagogical Feedback Framework for LLMs}
\subsection{Preliminary Interview}
To develop a feedback model for programming education and answer RQ1, we first conducted a preliminary interview with one computer science teacher (T3 in Tab. \ref{tab:demographics}). In a semi-structured interview, we asked the teacher to give feedback on a fictional prototype without functionalities (that we later turned into a feedback system, described below). During the course of 10 designed tasks on the topic of recursion, derived from computer science textbooks used in Germany, we asked the teacher how they would provide feedback to the presented task. Through this interview, we were able to identify three major adaptation criteria for feedback strategies used by the teacher: (1) Performance: Students with higher performance levels should receive less guidance. (2) Task Progress: Students who have not yet attempted to write code should be encouraged to do so, in order to promote active engagement and prevent a lack of student effort, and (3) Student Input: If students are unable to formulate questions independently, they should be offered guiding questions rather than direct assistance.
\subsection{Pedagogical Principles}

Based on the insights gained from the preliminary interview, and on prior studies, we first present key pedagogical principles we use in the framework before introducing the framework itself. The pedagogical principles differentiate the ideal feedback approach based on the student behavior: No coding attempt,  Code attempt that fails test cases, and Code attempt that passes test cases.

\subsubsection{(1) Pedagogical Principles for Students without Coding Attempt.}
The teacher described three strategies, which encourage students to start writing code and provide only high-level information that is not adapted to the specific mastery level promoting individual student effort \cite{narciss_how_2004}.
\begin{description}
    \item a) \textbf{Motivational messages} asking students for concrete questions are given when students seek help for the first time without a concrete question \cite{almeida_can_2012}. 
    \item b) \textbf{Guiding questions} are provided to promote self-directed learning and independent thinking \cite{boyer_principles_2010} when students cannot formulate their own questions. The more help they seek, the more specific and elaborative the questions should become \cite{mason_providing_2001}.
    \item c) \textbf{Targeted assistance} is offered when students have concrete questions or respond to responses generated before \cite{narciss_how_2004}. 
\end{description}
\subsubsection{(2) Pedagogical Principles for Students with Code Attempt Failing Test Cases.}
Pedagogical Principles for incompleted tasks depend on student mastery, which is an important adaptation criterion \cite{mason_providing_2001,narciss_how_2004}.
\begin{description}
    \item a) Low-performing students receive more explicit guidance by elaborative response-contingent feedback \cite{mason_providing_2001}, including partial code snippets. Additional focus should be on motivational support to reduce the tendency to give up \cite{gomes_paths_2015}.
    \item b) High-performing students benefit from broader hints and elaborative feedback on topics \cite{mason_providing_2001}. 
\end{description}
\subsubsection{(3) Pedagogical Principles for Students with Code Attempt passing Test Cases.} When a task is solved, feedback is provided with fewer restrictions, allowing the sharing of complete code snippets, as a solution has already been developed by the student. Responses are again adapted based on student mastery.
\begin{description}
    \item a) Low-performing students receive in-depth explanations of concepts to further help them understand the task and topic better.
    \item b) High-performing students receive encouragement to deepen their knowledge through explorative questions and answers. 
\end{description}
\subsection{Development and Implementation of LLM Feedback Framework}
\begin{figure}[t]
\centering
\includegraphics[width=\textwidth]{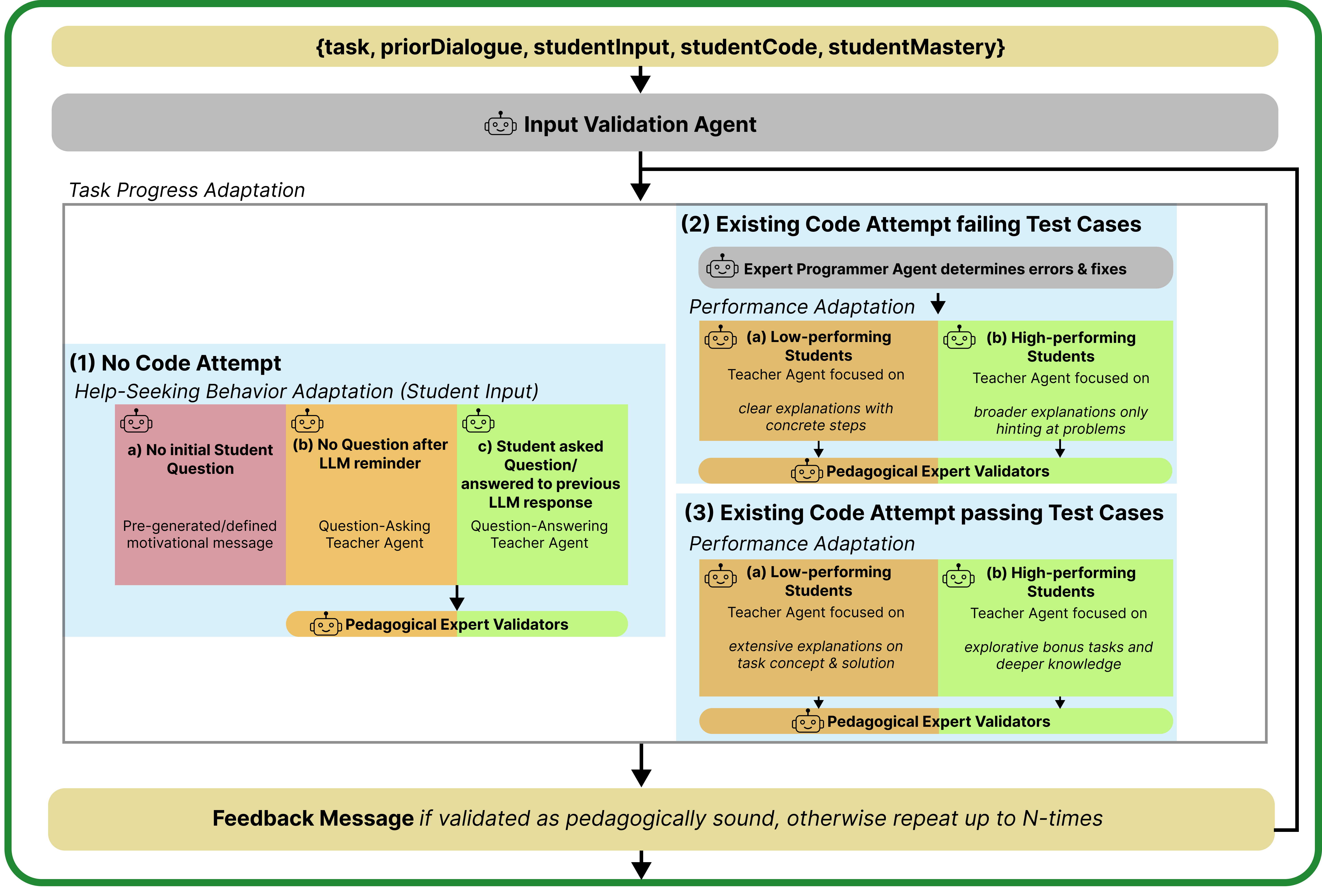}
\caption{Overview of our proposed Pedagogical LLM Feedback Framework for Programming Education. Each box (except the first and last) represents a single, unique prompt adapted to the specific scenario where the student is situated. A feedback message is returned successfully when enough validators assess them as pedagogically sound within a given number of iterations.} \label{fig:feedback Framework}
\vspace{-10pt}
\end{figure}

\looseness-1By synthesizing feedback strategies elicited from the teacher with previous feedback models \cite{mason_providing_2001,narciss_how_2004,nicol_formative_2006}, we propose a new framework for designing pedagogically-sound feedback using LLMs (see Fig. \ref{fig:feedback Framework}).  The framework proposes the following inputs to LLMs, enabling them to provide accurate, pedagogically-sound feedback: (1) Task Description, (2) Student Mastery Level, (3) Student Attempt, and (4) (Optional) Student Text Input, all of which have been addressed by the teacher from the interview. To facilitate a distribution of responsibilities for handling adaptive feedback messages' dynamic and complex nature, multi-agent LLM systems can be used \cite{PyTaskSyn-AIED2025,wang_llm-powered_2025,DBLP:journals/corr/abs-2308-08155}.

By employing different agents, instructed differently according to student profile, we can ensure adaptive responses promoting more individualized learning compared to existing LLM systems incorporating a single agent with a fixed prompt. To ensure students are not exploiting LLMs, we use a specific agent to determine whether this scenario occurs before attempting to give feedback. To prioritize the direct incorporation of pedagogical principles over direct source and target code analysis, Teacher LLM Agents for students with failing code attempts should avoid focusing on identifying errors in the student's source code themselves. Instead, we use an additional Expert Programmer LLM Agent to handle error analysis with the given sample code and test suite. Additionally, we use LLMs as Pedagogical Expert Validators after the feedback is generated by a specific teacher agent to verify the incorporation of specific adaptive pedagogical requirements.

\begin{figure}[t]
\centering
\includegraphics[width=0.8\textwidth]{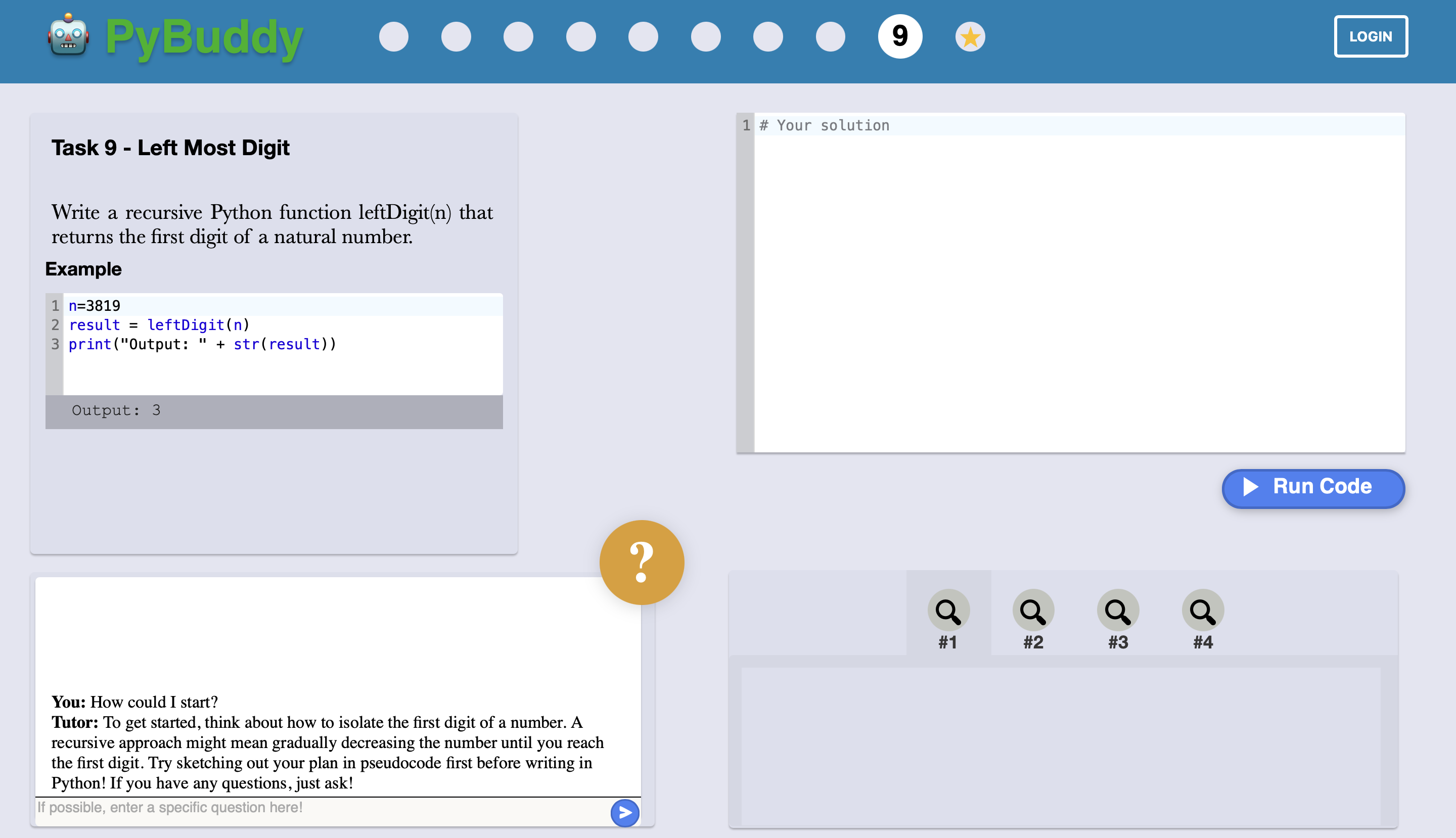}
\caption{Our web-based LLM feedback system (translated into English). Students can write code and run it against test cases on the right. On the bottom left, they can request help and get LLM-generated feedback.} \label{fig:evalPlatform}
\end{figure}

To evaluate LLM-generated feedback that aligns with our framework, we implemented a feedback system within a web application, including the same tasks as in the pre-interview, to support students' practice of recursion (see Fig. \ref{fig:evalPlatform}).

We used GPT-4o \cite{openai_gpt-4o_system_2024} and GPT-4o mini \cite{openai_gpt-4o-mini_2024} to implement the LLM Feedback. Parameters such as temperature %(see Tab. \ref{tab:agent_models}) 
The usage of 10 validation agents with seven approvals required for returning feedback within at most five iterations was determined through iterative testing.
We constructed prompts in German through iterative refinement based on our framework and pedagogical principles described in the previous section. Due to space constraints, we do not provide detailed prompts in this paper. German was chosen as the language for the tool since our evaluation was conducted in Germany and teachers taught programming to students in German. 
\section{Teacher Evaluation}
With our focus on instructing LLMs to provide pedagogically-sound feedback and to understand the extent to which LLM-generated feedback compares to human-teacher feedback, we conducted a mixed-methods evaluation with teachers through semi-structured interviews. To conduct the evaluation, we presented teachers with sample requests and corresponding LLM responses within our developed system.
\subsection{Participants}
We recruited eight computer science teachers (see Tab. \ref{tab:demographics}) within Germany (including one who provided initial insights in the preliminary interview). These teachers all had experience teaching programming at a secondary school in Germany and have taught Python (i.e., the content of our system). Teachers completed a consent form before participation and were compensated with 40€ for a 90-minute session. All interviews were conducted in German, and all teachers already had experience or knowledge of LLMs. Teachers chose to participate in the session either in person (T0, T3) or online (T1, T2, T4, T5, T6, T7). The study had received approval from the ethical review board at the authors' affiliation before data collection.
\begin{table}
\centering
\caption{Information about teachers participating in our study.}\label{tab:demographics}
\begin{tabular}{|l|l|p{4cm}|p{2cm}|p{1.8cm}|l|}
\hline
Teacher &  Gender & Federal State & Teaching & Teaching  & Evaluation \\ & & & Experience  & Grades & Group \\
\hline
T0  & Female & Saarland &  4 years &  3-12      & 1 \\
    T1  & Male   & Rhineland-Palatinate &  20 years & 5, 6, 10-13 & 1   \\
    T2  & Male   & North Rhine-Westphalia &  20 years & 5-13       & 2   \\
    T3  & Male   & Saarland &   2 years &  7-10, 12  & 2  \\
    T4  & Male   & Rhineland-Palatinate &  18 years & 9-13       & 3  \\
    T5  & Male   & Hamburg &  4.5 years & 5-13       & 3  \\
    T6  & Male   & Saarland &  8 years & 7-12       & 1   \\
    T7  & Male   & North Rhine-Westphalia & 2 years & 7-13       & 3  \\
\hline
\end{tabular}
\end{table}
\subsection{Evaluation Data}
Our evaluation was designed to test the effectiveness of the framework and system on whether it can provide pedagogically-sound feedback. To simulate student work that our system can use to generate feedback, we manually created synthetic student code and requests based on common error patterns and scenarios for recursion tasks identified in our preliminary interview. Also, we used GPT-4o \cite{openai_gpt-4o_system_2024} to generate controlled variations of the questions and source code on all 10 tasks within our system. All generated variations were manually reviewed. If they contained overly formal phrasing or an unnatural code style, we adjusted them to better reflect realistic student language and coding style.

\begin{figure}
\centering
\includegraphics[width=0.85\textwidth]{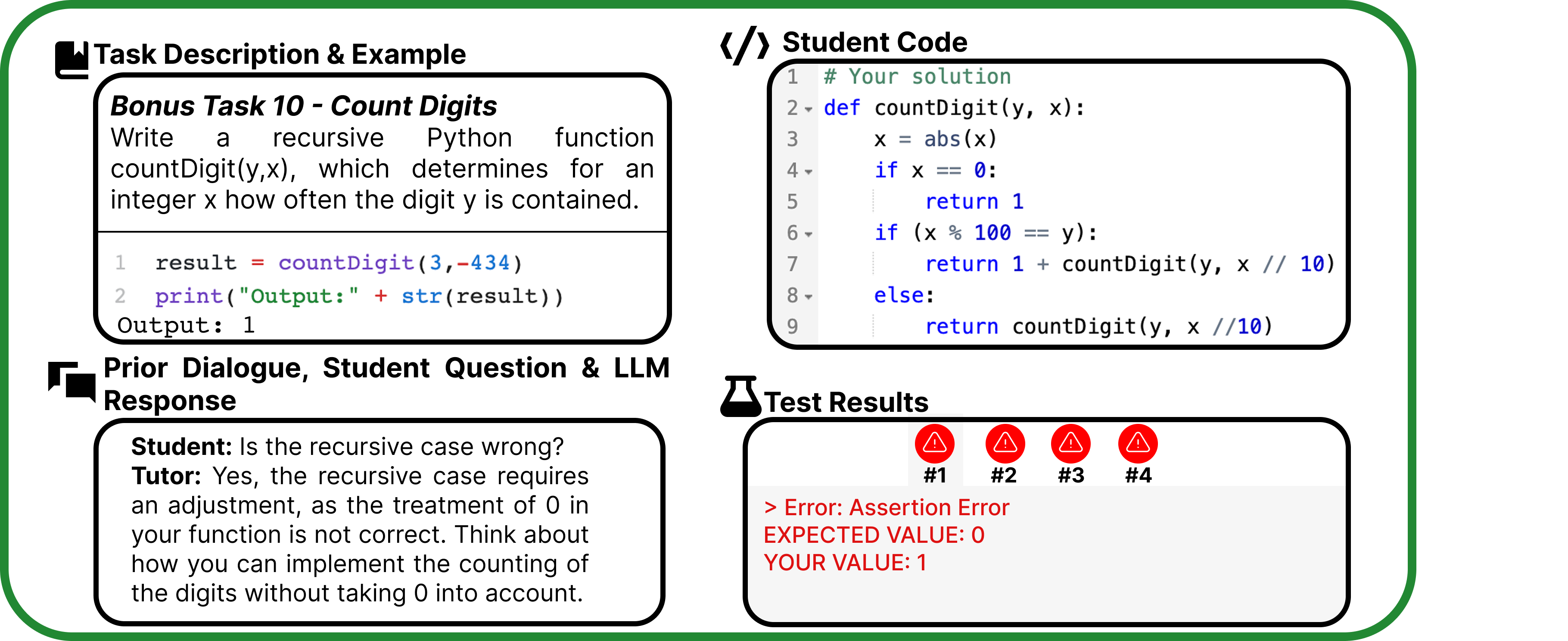}
\caption{An exemplary artifical student request displaying a common mistake of making small errors within the recursive case, translated to English.} \label{fig:StudentRequest}
\end{figure}

\looseness-1One such artificial student help request in our mock dataset includes source code (and corresponding test cases, if available), text input, and mastery level (see Fig. \ref{fig:StudentRequest} for an example). We pre-generated responses from our LLM system and attached them to the dataset. To reduce potential bias in generated evaluation requests, we did not create the exact number of requests and responses needed for teacher sessions, but instead a larger mock dataset of 329 student requests and system responses covering all 10 recursion-related tasks, with varying scenarios identified during our preliminary interview. Due to time constraints, no more than 20 pre-generated responses could be evaluated per session. To ensure broader qualitative and quantitative insights across diverse scenarios, we randomly sampled 60 requests from the dataset, evenly distributed across three distinct evaluation groups (see Table \ref{tab:demographics} for group assignments). Each group included different study participants, student requests, and corresponding LLM responses. Our analysis accounted for potential variation in scenarios discussed by different teachers.
\subsection{Procedure}
We conducted semi-structured interviews that were structured into five parts. After some warm-up questions on demographics, we asked the educators about their teaching practices and current use of LLMs. We then asked how teachers define good feedback to better interpret their ratings in analysis. Next, we introduced teachers to our system (see Fig. \ref{fig:evalPlatform}). They were asked to try out, ask for feedback (as if they were a student), and explore the system to better understand how LLM feedback systems work and approximate feedback generation time. Then, we asked teachers to evaluate 20 student requests, also in terms of their realism, and the corresponding LLM responses within our system. Help requests were labeled with corresponding mastery level attributed to the scenario based on insights from the pre-liminary interview (Weak, Strong, or No Coding Attempt). Each teacher was randomly assigned one of the three evaluation groups before the interview, each of which contained different student requests and LLM responses to get broader insights on different scenarios.

For each student's request, we explicitly asked teachers how they would give feedback without revealing the LLM response to understand the differences between LLM and teacher responses with reduced bias. Then, the LLM response was revealed, and we discussed it together with the study participant. To facilitate discussions and gather additional quantitative data, we asked teachers to answer six questions in our system using a Likert, Ternary, or Binary scale (see Tab. \ref{tab:questionsEval}). After the evaluation, we engaged teachers in a reflection session about their views on our feedback system and potential improvements. We also asked them to compare it to popular tools such as ChatGPT.
\begin{table}[!b]
\centering
\caption{Questions and Scales of Questions within our Evaluation Dashboard.}\label{tab:questionsEval}
\begin{tabular}{|l|l|p{6cm}|}
\hline
    \textbf{Metric} & \textbf{Scale} & \textbf{Additional Explanations} \\ \hline
    {\textit{Correctness}} & Yes / Partially / No & Teachers were asked whether the response was technically correct \\  
     {\textit{Pedagogically Sound}} & 5-Point Likert Scale & Teachers were asked whether it adheres to their understanding of pedagogical feedback. \\
     {\textit{Comprehensive}} & 5-Point Likert Scale & Teachers were asked if students would comprehend the answer. \\   
     {\textit{Effective}} & 5-Point Likert Scale & Teachers were asked whether the response helps low/high-performers progress. \\   
     {\textit{Worse/Better than Own Feedback}} & 5-Point Likert Scale & Teachers were asked to compare the LLM response with their usual feedback in the scenario. \\    
    {\textit{Need for Edits}} & Yes / No & Teachers were asked whether they would change something about the LLM response to facilitate discussions about differences. \\
    \hline
  \end{tabular}
\end{table}

\subsection{Analysis}
\subsubsection{Qualitative Analysis}
We transcribed and inductively coded the interviews, with the goal of evaluating and comparing LLM-generated feedback with human teacher feedback.  Before analysis, we first structured our data (quotes + interpretative codes) into five sections: Pre-Evaluation Comments, Mid-Evaluation (General Comments, Differences among Teachers on same Help Request, Similarities among Teachers on same Help Request), and Post-Evaluation Notes.

We used Affinity Diagramming \cite{krause_affinity_2024} to obtain qualitative insights, a standard analysis technique in Human-Computer Interaction that allows the formation of clusters of higher-level themes within qualitative data. Both quotes and codes contributed to the formation of clusters to reduce potential bias during coding. First, we performed Affinity Diagramming on the codes within each of the five sections, resulting in 120 low-level themes altogether. %(17 + 21 + 30 + 35 + 17). 
Then, we performed Affinity Diagramming across the five cluster groups to formulate high-level themes, which resulted in 20 high-level qualitative themes (that contained the above-mentioned 120 low-level themes).
\subsubsection{Quantitative Analysis}
As outlined in the previous section, we asked teachers to rate feedback messages. Those ratings on six questions (see Tab. \ref{tab:questionsEval}) were used to calculate means (and SDs) of the ratings across teachers. We also calculated the mean for individual requests to get findings about average perceptions of individual LLM responses. 

Further, during the qualitative coding, we observed that there were some disagreements among different teachers on their perception of the LLM-generated feedback. To support this qualitative disagreement quantitatively, we calculated disagreement scores following Whitworth's formula \cite{whitworth_measuring_2007}, which measures disagreement on a scale of 0 to 1. The higher the score, the higher the disagreement. To ensure the accurate assessment of disagreeing perceptions with disagreement scores, we transformed our 5-point Likert Scale Answers into a scale with three points (1 \& 2 to negative, 3 to neutral, 4 \& 5 to positive). 
\section{Results}
\subsection{Unique Benefits and Complementing Support by LLM-Generated Feedback}
\begin{figure}
\includegraphics[width=\textwidth]{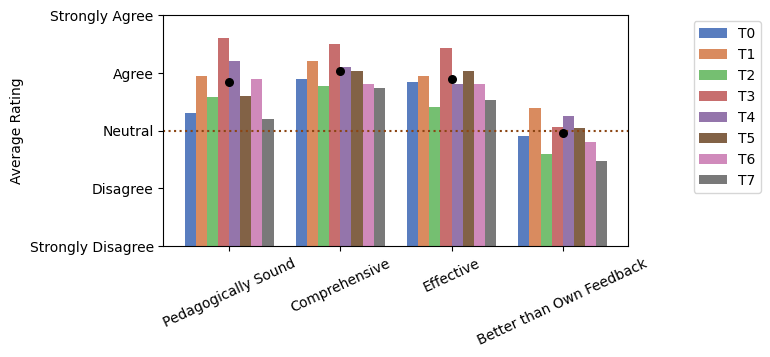}
\caption{Average Pedagogical Soundness, Comprehension, Effectiveness and Comparison Ratings by participant (single bar) and overall (black dot).} \label{fig:barcharts}
\end{figure}

We found from the quantitative data that all teachers generally considered that LLM-generated feedback is well aligned with our developed framework, and it is pedagogically sound (Mean: 3.84, SD: 1.10), highly comprehensive (Mean: 4.04, SD: 1.01), and effective in supporting student learning (Mean: 3.89, SD: 1.14) (see Fig. \ref{fig:barcharts}). Our qualitative analysis showed two main benefits of LLM Feedback.

\subsubsection{LLMs offer broader and deeper knowledge for student support with respect to their mastery level.} Teachers appreciated our LLM System, and LLMs in general, for their deeper knowledge, supporting high-performing students (not only low-performing students) through providing further support to deepen their knowledge. For example, one teacher mentioned for an explorative request of a high performer: "I would not have any idea how the student could extend [the task] [...]. That's what the AI is really good at" (T4). One teacher explained: "I would need to spend more time for lower-performing students" (T5), which means that LLM Systems can uniquely support high-performing students, while teachers can use their (limited) time for low-performing students.
 
Supporting this insight, our LLM feedback for high performers received the highest average ratings for pedagogical soundness (Mean: 4.02, SD: 0.63) and effectiveness (Mean: 4.14, SD: 0.68), while receiving slightly lower scores than low-performers on understandability, likely because of more complex explanations involved in the cases for higher-performing students (see Tab. \ref{tab:scoresMastery}).

Teachers consistently emphasized that LLM-generated feedback is comprehensive and pedagogically sound, as supported by quantitative scores (see Fig. \ref{fig:barcharts}). They appreciated the goal-oriented nature of our LLM system, which was instructed to provide concrete steps to help the student close the gap and solve the task. They expressed that they liked the LLM focusing its feedback on correcting fundamental issues before simply answering other minor questions, even though it may initially confuse students (T2). %Teachers also liked our adaptability mechanisms, explicitly stating that the specific LLM response can support strong students (T2, T4), while others specifically support weak students (T1, T2, T4). 
\begin{table}[h]
\caption{Mean LLM response scores by student mastery.}\label{tab:scoresMastery}
\centering
\begin{tabular}{|l|l|l|l|}
\hline
   & \textbf{High-performing} & \textbf{Low-performing} & \textbf{No Code Written} \\ \hline
        {Pedagogically Sound} & 4.02 (SD: 0.63) & 3.84 (SD: 0.82)  & 3.76 (SD: 0.58) \\ 
        {Comprehensive} & 4.00 (SD: 0.88) & 4.17 (SD: 0.44) & 3.89 (SD: 0.72) \\  
        {Effective} & 4.14 (SD: 0.68) & 3.98 (SD: 0.71) & 3.52 (SD: 0.88) \\  \hline
        Rated Requests & 17 & 21 & 22 \\
    \hline
  \end{tabular}
\end{table}
%\vspace{-20pt}
\subsubsection{LLMs excel in Error Detection.} Many study participants expressed that our LLM system would be significantly faster and more effective in identifying small errors in student code, such as syntax errors or missing calculations. One teacher stated, "The AI would probably be way better than a teacher [...]. I might have had to look at this code for 5 minutes because it takes time to find such errors" (T2). All teachers rated the provided LLM responses as highly accurate (91.7\% correct (55), 8.3\% partially correct (5)).
In particular, teachers appreciated that the answers of our LLM System were "more precise than verbal feedback that [they] usually give" (T1). 

\subsection{Differences in Understanding of Pedagogically Sound Feedback}
Despite the findings on unique advantages that LLMs provide in feedback giving in programming education, we also observed critical discrepancies among teachers in their ratings and perceptions towards the feedback generated within the system. For example, on the pedagogical soundness, the ratings ranged from 3.20 (T7) to 4.61 (T3). Such a difference indicates that our LLM System and Framework might not be able to meet all teachers' desires.

Indeed, our calculated disagreement scores (see Tab. \ref{tab:disagreement scores}) on Pedagogical Soundness and Comparison support those subjective perspectives on pedagogical principles and the challenge of designing feedback systems that satisfy all expectations of teachers. 
\begin{table}
\caption{Disagreement Scores for all Groups on Pedagogical Soundness and Comparison Ratings.}\label{tab:disagreement scores}
\centering
\begin{tabular}{|l|l|l|l|}
\hline
\textbf{Question} &  \textbf{Group 1} & \textbf{Group 2} & \textbf{Group 3} \\
\hline
Pedagogically Sound & 0.4000 & 0.5000 & 0.4833 \\
Better/Worse than Own Feedback & 0.7000 & 0.5500 & 0.7500 \\
\hline
\end{tabular}
\end{table}
To further understand the disagreements among teachers, we searched within our high-level and low-level themes, finding two main factors that contribute to the disagreement: (1) \textbf{Different perceptions towards appropriate student language}, specifically whether technical terms, such as "Stack Overflow" were considered too advanced (T3) or not (T2). (2) \textbf{Varying views on Hint Specificity}: While some teachers thought that fine details, such as revealing missing calculations (T6), were effective and helpful, others did not think so (T1).

\subsection{Limitations of LLM-generated Feedback}
While teachers generally considered our LLM-generated feedback pedagogically sound, our qualitative analysis highlights three key limitations of LLM Feedback: 
\subsubsection{LLMs struggle to adapt to classroom context and student needs.} Unlike teachers, who dynamically adapt their feedback to current lessons and students' individual characteristics, our LLM Framework and system is unable to tailor its response to specific situational contexts. One teacher explained, "It totally depends on where you are in your lesson" (T4), supported by another stating: "I would only answer this question if we've discussed it in class already" (T5). Such contextual knowledge allows teachers to provide more relevant and personalized feedback, which, without the incorporation of individual teacher instructions to the LLM, general LLM Feedback Systems would not be able to provide. 

\subsubsection{LLMs lack interactive, real-time interaction and engagement with students.} Teachers believe that they have a better understanding of students' thought processes, as they can see the documentation of their progress on paper that the LLM cannot see, which allows them to intervene during problem-solving when detecting errors (T5). Typical LLM Systems, such as ours, give students full control over when to receive feedback, which can result in missed opportunities for quick interventions. 
\subsubsection{LLMs lack interactive and visual teaching methods.} Teachers in our study proposed various strategies for supporting students, which LLMs cannot use. Teachers prefer helping students by using additional materials, such as a sheet of paper or a blackboard, to illustrate concepts visually. One teacher said, "I would do an example on a sheet of paper; I like working on a sheet of paper so they understand the task" (T5), while another adressed the importance of visualizing concepts, especially for struggling students: "I would ask the student to write an example down visually. This helps a lot with comprehension" (T3). Text-based LLM Feedback systems like ours cannot provide the same visual learning experience as teachers.
\section{Discussion}

In this work, we presented a new framework for providing pedagogically-sound feedback in programming education and tested its approach through implementing and evaluating a feedback system with computer science teachers. Our results add important implications on teachers' desires for more adaptive, pedagogically-aligned feedback systems imitating teacher feedback strategies. 

Although prior studies already recognized the capabilities of LLMs for adaptive feedback and explored how to provide accurate and efficient LLM-generated feedback, they did not keep pedagogical principles in mind \cite{ma_how_2024,phung_automating_2024}. To address this gap and our first design/research question, we conducted a preliminary interview and developed a framework that focuses on pedagogical principles. Our findings indicate that these pedagogical principles, when implemented as adaptive prompts instead of fixed prompts, were positively evaluated by teachers.

Regarding our second research question on the extent to which LLM-generated feedback compared to human teachers provides valuable feedback, teachers have praised LLMs for their ability to handle routine queries in parallel, which can help overcome the challenge of a larger class size. However, the study also highlights critical limitations of LLMs, underscoring the importance of human teachers in classroom teaching.
In particular, we showed that the lack of real-time interactions and contextual adaptability reinforces the idea that LLMs can augment but not fully replace human-generated feedback. We demonstrated that teachers' feedback practices are highly individualized and contextualized, making a common application of LLMs impractical as existing work tends to take a one-size-fits-all LLM approach. While LLMs hold high promise in reducing teacher workload, their effectiveness highly depends on individual teachers and how much the feedback could be configured to properly align with classroom dynamics, which needs to be acknowledged when developing feedback systems.
\subsection{Limitations}
We acknowledge that our study has some limitations. One key limitation of our study is the potential biases of the teachers participating in our study, which could have influenced the evaluation. Prior experience with LLM platforms, including ChatGPT, may have shaped their ratings and qualitative input. 

Further, the external generalizability of our findings is constrained by the study's small sample of eight computer science teachers in Germany. While our participants varied in experience, they might not represent a broader population of computer science teachers. Another concern is the construction of the dataset, as the student prompts and code snippets were designed by researchers instead of being drawn from the real-world classroom setting due to time constraints.

\section{Conclusion \& Future Work}
Our work offers a vision for future automated feedback systems to better reflect real classroom practices, support student diversity, and respect teacher autonomy in terms of giving feedback by placing pedagogical principles at the center of system design. Our findings show that LLMs can be an important companion to teachers when it comes to giving feedback in the classroom (e.g., based on their capabilities of adapting to mastery levels and the advantage of parallel support). However, significant limitations mentioned by teachers during evaluation show that they lack variability when it comes to adapting to diverse feedback types used by individual teachers. Hence, from the teachers' perspective, they should not continue to be a one-size-fits-all system. This marks a shift from viewing LLMs as static tools for classroom usage to adaptable partners cooperating with both teachers and students.

Future work will be essential to further extend our findings, such as an evaluation with students to investigate whether the focus on pedagogical principles within prompts increases feedback efficiency. Additionally, the refinement of adaptation mechanisms within our framework (e.g., by richer student modeling or shift to different domains) could be investigated to further enhance the adaptiveness of Automated Feedback Systems tailored to individual students in various domains. Building on research suggesting adaptive feedback \cite{mejeh_effects_2024}, our study suggests that multi-agent LLM systems, rather than single-prompt systems, offer a more flexible way to tailor feedback based on student mastery levels and task progress.

\subsubsection*{Acknowledgments} This work was partially supported by JST, PRESTO Grant Number JPMJPR23I8, Japan. We thank all the teachers participating in our study.
\bibliographystyle{unsrt}  
\bibliography{references}  %%% Remove comment to use the external .bib file (using bibtex).


\end{document}